\documentclass[conference]{IEEEtran}
\IEEEoverridecommandlockouts
\usepackage{cite}
\usepackage{amsmath,amssymb,amsfonts}
\usepackage{algorithmic}
\usepackage{graphicx}
\usepackage{textcomp}
\usepackage{xcolor}
\def\BibTeX{{\rm B\kern-.05em{\sc i\kern-.025em b}\kern-.08em
    T\kern-.1667em\lower.7ex\hbox{E}\kern-.125emX}}

\usepackage{xurl}
\usepackage{hyperref}

\usepackage{fancyhdr}
\begin{document}

\title{Grounding Insight: An Artificial Affect Signal for Human-Centric Explainability of RL Agents}

\author{\IEEEauthorblockN{1\textsuperscript{st} Bernhard Hilpert}
\IEEEauthorblockA{\textit{LIACS} \\
\textit{Leiden University}\\
Leiden, Netherlands \\
0000-0002-0011-2905}
\and
\IEEEauthorblockN{2\textsuperscript{nd} Kim Baraka}
\IEEEauthorblockA{\textit{Department of Computer Science} \\
\textit{Vrije Univeriteit (VU) Amsterdam}\\
Amsterdam, Netherlands \\
k.baraka@vu.nl}
\and
\IEEEauthorblockN{3\textsuperscript{rd} Joost Broekens}
\IEEEauthorblockA{\textit{LIACS} \\
\textit{Leiden University}\\
Leiden, Netherlands \\
joost.broekens@gmail.com}
}

\maketitle
\thispagestyle{fancy}

\begin{abstract}
In Human-Interactive Robot Learning (HIRL), human teachers assess learning progress by observing agent behavior. 
However, Reinforcement Learning (RL) agents often fail to express learning progress in human-interpretable ways, leading to suboptimal or counterproductive teaching. 
This work proposes ``Insight": a human-interpretable affective signal grounded in the RL process. It is specifically designed to address the ”Absence of learning” human teacher mental model misalignment - the misconception that a failing or exploring agent is broken, underpowered or not learning at all - by signaling an agent’s ”Aha!” Insight moment. The signal is derived from the absolute Temporal Difference (TD) error making it easily implementable in other RL agents. To ground this signal, five design criteria were formulated: Relevance, Interpretability, Faithfulness, Scalability, Generalizability. The structural construct validity was assessed through simulation, examining micro and macro dynamics and event-related sensitivity. Micro dynamics reveal that Insight captures meaningful learning events at the appropriate time and magnitude. Macro dynamics demonstrate that the signal follows the agent’s learning phases and reflects overall progress. Event-related sensitivity shows, the signal distinguishes the saliency of environment interactions, such as positive rewards versus ``trauma-like" negative events. These results demonstrate that technical RL dynamics can be effectively packaged into human-like affective substrates to close the teacher-learner loop.
\end{abstract}


\section{Introduction}
In Human-Interactive Robot Learning (HIRL) \cite{baraka_humaninteractiverobotlearning_2026}, human teachers are integrated into the robot learning process through interactive feedback cycles to assist robot learners on adaptively learning a task they otherwise would not have been able to learn on their own \cite{akataResearchAgendaHybrid2020}. 
Reinforcement Learning (RL) as a formalism is suitable for sequential decision-making under uncertainty and is therefore very applicable to robots that learn through interaction with their environment and potentially human teachers, the focus of this paper.

However, in practice these settings often suffer from teacher-learner mental model misalignments \cite{richterImprovingHumanRobotTeaching2025}: while the robot learns from the human teaching signals, the process of learning is a ``black box" and not intuitively understandable to human teachers \cite{hellstromUnderstandableRobotsWhat2018}, especially if they are domain- but not machine learning experts \cite{habibianReviewCommunicatingRobot2023}.
In such cases, the actual effect of the teaching signals remains opaque to most teachers themselves \cite{nguyen_creditassignmentchallenges_2023} and often times they can can be unsure about whether and how the agent is learning in general \cite{tabrezSurveyMentalModeling2020, hilpertCanYouSee2025}. 
While some post-hoc xRL methods have attempted to compute agent-centric metrics like `interestingness' to highlight critical states \cite{sequeiraInterestingnessElementsExplainable2020}, the underlying algorithmic formalisms remain concepts designed for RL experts rather than non-expert human teachers.
Consequently, they provide technical transparency, but lack the evaluative quality necessary for alignment with a human's mental model. For instance, a teacher observing an agent repeatedly failing a task (or repeating behavior in general) might conclude that no progress is being made (``Absence of Learning"). This may wrongfully trigger a teacher to instruct the agent to try something else. Yet, behind this lack of visible progress, the agent's internal update rule may be successfully integrating vital reward information. Internal ``Insight" remains invisible to the observer until it suddenly manifests as behavioral change. This demonstrates that more human-friendly, intuitively interpretable signals are needed.

In biological agents, affect plays an important role as an internal evaluative currency \cite{feldman_neurobiologyinteroceptionaffect_2024}. It acts as a mechanism for internal disambiguation, allowing an agent to assess the relevance of new information relative to its goals \cite{moors2019demystifying, schillerHumanAffectome2024}. 
Beyond its internal function, affect is fundamentally integrated into social interaction; humans express internal affective states, and, intuitively make sense of the behavior of others by attributing assumed affect. 
In the context of RL
, simulating affective states may provide a structured way to summarize the agent's functional internal learning status that can serve a dual purpose: 
 it faithfully reflects the complex internal dynamics of RL processes, while simultaneously providing a basis for expressive cues that humans are predisposed to intuitively interpret. 
Affect has been simulated from RL (e.g., \cite{moerlandEmotionReinforcementLearning2018, broekensTemporalDifferenceReinforcement2018, el-nasrFLAMEFuzzyLogic2000}), but has not yet been deployed as an explainability mechanism to bridge mental model misalignment in HIRL settings.

To effectively support teacher alignment, 
artificial affect must faithfully reflect the teaching-relevant aspects of an agent's actual learning process.
We therefore formulate five specific design criteria and introduce a rigorous four-stage framework for deriving and validating RL-grounded affective signals: Empirical Grounding, Signal Conceptualization, Algorithmic Implementation, and finally Multi-Criterion Validation.

By following this framework, our primary contributions are:
\begin{enumerate}
\item
Five design criteria (Relevance, Interpretability, Faithfulness, Scalability, and Generalizability) to guide the synthesis of process-reflective artificial affect,
\item
The ``Insight" signal, an algorithm-faithful, human-interpretable explainability mechanism derived from the TD-error, specifically designed to address the ``Absence of learning" mental model misalignment by signaling an agent's ``Aha!" moments,
\item
A simulation-based evaluation showing that Insight successfully disambiguates learning from habitual execution
\end{enumerate}
This ``proof-of-concept" establishes a blueprint for using affective substrates to enhance explainability in HIRL.

\section{Background}

HIRL relies on the teacher-learner loop in which the teachers give feedback signals based directly on the learning behavior the robot exhibits \cite{baraka_humaninteractiverobotlearning_2026}.
This often times relies on implicit communication, where teachers infer the robot's internal learning progress through its external behavior 
 \cite{habibianReviewCommunicatingRobot2023}.
Previous research suggests, that teachers apply anthropomorphic frameworks for these inferences, projecting human-like concepts, like ``knowledge" and ``learning mechanisms" onto RL agents \cite{hilpertCanYouSee2025}.
Consequently, teachers adopt diverse pedagogical roles and teaching strategies based on these subjective assumptions rather than the agent's actual state \cite{hilpert2026toss}.  
Because adaptive agents learn through algorithms that differ significantly from human learning, a fundamental misalignment arises between teacher assumptions and the actual algorithmic formalisms \cite{richterImprovingHumanRobotTeaching2025}.
Thus, human mental model misalignment is not merely a passive misunderstanding but is highly consequential. 

To realign the human mental model without increasing the teacher's cognitive load, an explainability signal must be both algorithmically faithful and intuitively interpretable. 

In biological systems, affect serves this exact dual purpose \cite{schillerHumanAffectome2024}.
Affect is ingrained in a wide range of mental processes \cite{dukesRiseAffectivism2021}, from perception \cite{wentura_meddlinginaffectiveinformation_2003}, interpretation \cite{barrett_theoryconstructedemotion_2017}, thoughts, over memory up to action tendencies \cite{lowe_feelingactiontendencies_2011}, motivation \cite{cardinal2002emotion}, behavioral expression \cite{keltner_emotionalexpressionadvances_2019} and social interaction \cite{fischerSocialFunctionsEmotion2016}. 
Functionally, affective processes act as an internal ``common mental currency" that evaluates ongoing events in relation to an organism's current goals \cite{moors2019demystifying, maturanaAutopoiesisCognitionRealization1991, moors2017integrated, peters2006functions}
, assigning relevance to competing interpretations \cite{barrett_theoryconstructedemotion_2017} and action options 
\cite{feldman_neurobiologyinteroceptionaffect_2024} at different levels of abstraction. 
Thereby, internal affective signals represent a continuous, first-person perspective on the underlying evaluation of mental processes.
In a learning setting, this translates to, e.g., evaluating progress within a sequential decision-making task \cite{sequeira2011emotion}.

Importantly, this disambiguating function of affect extends beyond internal evaluation to the social disambiguation of behavior \cite{fischerSocialFunctionsEmotion2016}. 
In interactive settings, humans naturally display behavioral expressions of these internal assessments, making it accessible to an external observer.
Attributing such affective states is an inherent component of how observers intuitively make sense of others' behavior \cite{scarantino_emotionalexpressionsappeals_2022, schneebergerCanSocialAgents2019}. Thereby, communicative affective signals complement, rather than interrupt an observer's inferences and serve as a social currency that supports mutual understanding in an intuitive way. 
In a teacher-learner setting, this means the teacher can use these affective cues to intuitively contextualize the learner's behavior and adapt their teaching based on this 
\cite{d2012dynamics, d2014confusion}.
Multidisciplinary research \cite{dukesRiseAffectivism2021, schillerHumanAffectome2024, scherer_handbookaffectivesciences_2009, calvo_oxfordhandbookaffective_2015, lugrin_handbooksociallyinteractive_2022} establishes affect as a bridging mechanism because it is simultaneously ingrained in process execution and inherently human-understandable.

Previous work has addressed this idea for Human-Agent Interaction (e.g. \cite{demeloReadingPeoplesMinds2014, loweGroundingEmotionsRobots2016}) and even RL (\cite{broekensEmotion2021, broekensTransparentRobotLearning2019, hilpert2024closing}).
As surveyed by Moerland et al. \cite{moerlandEmotionReinforcementLearning2018}, emotions can be systematically derived from an agent's decision-making architecture, serving to steer action selection and communicate internal states. 
More recently, Barthet et al. \cite{barthet_affectivereinforcementlearning_2026} provided a comprehensive taxonomy of affective reinforcement learning, laying out how affective information is interwoven within RL to shape rewards, drive action policies, and form part of the state representation. 
A prominent framework within this space is the Temporal Difference Reinforcement Learning (TDRL) theory of emotion \cite{broekensTemporalDifferenceReinforcement2018}, which models affective intensities as manifestations of temporal difference (TD) error signals. This approach offers theoretical continuity between artificial agents and biological organisms by deriving simulated internal evaluations from the agent's actual learning progress. 
Previous research has demonstrated that such RL-grounded signals yield psychologically plausible intensities \cite{daiSimulatingFearAnticipation2021, broekensTDRLModelEmotion2019, moerlandFearHopeEmerge2016, zhang2024simulating, barros_incorporatingrivalryreinforcement_2023, chen_trustwesurvive_2026} and can be externalized as believable behavioral expressions \cite{matareseTransparencyTDRLRobotic2020, anijeholtRoleSimulatedEmotions2023}.

While these models have proven successful at improving learning efficiency and facilitating social believability, their potential as a human-centric explainability mechanism in active teaching scenarios remains largely unexplored. In HIRL, affective signals must reflect the learning process itself to allow human teachers to contextualize behavior and adapt their interventions. By serving as a continuous, naturally interpretable ``post-hoc saliency method" \cite{habibianReviewCommunicatingRobot2023}, an RL-grounded affective signal provides a shared currency to help observers infer why an agent is acting the way it is acting, thereby building the basis for reducing mental model misalignments and improving instructional efficacy. 
Formalizing RL-internal dynamics as such an affective signal can leverage the natural human intuition to better align the teacher’s mental model with the robot's actual learning progress in real-time.
To effectively support this alignment in interactive teaching settings, an affective signal must fulfill a critical prerequisite: it must provide an internally grounded evaluative signal that faithfully reflects the teaching-relevant aspects of an agent’s actual learning process.
This work therefore formalizes a process-reflective affective signal within an RL agent, demonstrating how internal learning mechanics can be translated into continuous, evaluative signals designed for human-centric alignment.

\section{``Insight" Model}

\subsection{Design Criteria}
In order to conceptualize specific affective signals for the HIRL context, five design criteria were specified to guide the formalization, inspired by  \cite{hudlickaHabitsStandardsSystematic2014}:

\begin{enumerate}
    \item Relevance (Teaching Focus): An artificial affective signal should address an empirically identifiable mental model misalignment.
    \item Interpretability: An artificial affective signal should be mappable to an intuitively human-interpretable concept to tap into humans' natural attribution framework and minimize the cognitive load of the external observer.
    \item Faithfulness: An artificial affect signal should reflect (align with) a relevant underlying internal evaluative mechanism of the RL process.
    \item Scalability: The formalization of affective signals should capture the temporal dynamics of the relevant RL process, irrespective of temporal scale or granularity, allowing task- and context-independent (scalable) simulation.
    \item Generalizability: An artificial affect signal should be derived from standard evaluative architectural components of the RL formalism to ensure broad applicability across diverse RL architectures.
\end{enumerate}


\subsection{Relevance: Mental Model Misalignment}
To identify teaching-relevant misalignments, this work examined an existing dataset of human inferences about RL agents and their functioning 
 \cite{hilpertCanYouSee2025, hilpert2026toss}, that is 
 openly accessible on the OSF and contains 816 statements of how observers make sense of RL agents' learning processes \cite{hilpertCanYouSee2025} and 204 teaching intentions \cite{hilpert2026toss}. 
To satisfy the first criterion and pinpoint a specific, teaching-relevant misalignment, the individual qualitative statements were systematically analyzed.
An important identifiable misalignment concerns the perception of 
learning progress
, characterized by three distinct patterns\footnote{This section represents a summary of this analysis. For an extensive description, please refer to the additional material available on the OSF}:
\begin{enumerate} 
    \item Observers often \textbf{mistake exploration for randomness}, ``goallessness" 
    or simply ``no learning", assuming that the agent is not capable of goal directed learning. This can be extracted from statements such as \textit{``The robot is not learning here at all"}, ``\textit{It didn't learn anything}" or “\textit{The robot seems to be making its decisions randomly}”.
    \item Observers often \textbf{overlook gradual improvements} and underestimate the time needed for convergence or backpropagation, essentially acknowledging that learning exists but does not function correctly or not how they would expect it. This can be seen in statements such as \textit{``Over the three attempts, it clearly gains knowledge, That knowledge then vanishes as the robot make another error on its third attempt}” and \textit{``An accidental movement reached the goal"}.
    \item Observers often \textbf{infer misprioritization of subgoals} in the agent with corresponding overfitting of the learning process, essentially assuming a learning process that is less powerful than the RL mechanisms actually are. This pattern can be seen in statements such as \textit{``Avoiding furniture is clearly a primary goal, but… gets stuck when furniture is in the way”}
    and \textit{``It's not really making decisions, just kinda doing stuff until something works.”}.
\end{enumerate}

Overall, human observers seem to equate visible task success with learning, while the actual state of the learning process is not always visible in RL agent behavior. Agents often learn most when they appear to fail as visible changes in action selection due to backpropragation of values takes time, and depend, e.g., on the action selection mechanism.
This misalignment can lead to a skewed teacher mental model of the learning agent's functioning, particularly in terms of understanding the granularity and dynamics of the (not always explicitly visible) learning process. 
Such a misalignment is especially crucial in real-time teacher learner interactions.

\subsection{Signal Conceptualization: Insight as an Affective Signal}

To bridge this teaching-relevant (Criterion 1) misalignment between visible success and internal progress, the signal must provide a human-understandable reflection of actual learning.
This work proposes an artificial affect signal, grounded in the RL learning dynamics (\textit{Faithfulness}), while \textit{remaining accessible} to intuitive human interpretation (\textit{Interpretability}).
The signal should capture moments in which the agent has learned something \textit{meaningful}: instances of learning that produce \textit{relevant internal change} within the agent’s value estimations, policies, or state–action associations.
Interpretability is facilitated by anchoring the signal to the human equivalent of an ``Aha''-moment in human learning, defined as the sudden, salient recognition of a solution following a period of impasse \cite{bowden2003aha, kounios2009aha}.
By adopting this human equivalent as a conceptual metaphor for the artificial affect signal, it relates to the intuition that: \textit{``I (the agent) have learned something meaningful here."}. Conceptually that can be summarized as \textbf{Insight}.

\textbf{Definition of Insight (in Human-Interactive Robot Learning):}
\textit{Insight is a process-reflective affective signal, mathematically grounded in the RL process, that captures meaningful internal learning events in a human-understandable conceptual frame. Its function is to make hidden learning dynamics transparent, specifically signaling when an agent has integrated feedback from experience and thereby proactively correcting teacher misconceptions that exploration or failure imply an ``absence of learning". Ultimately, Insight serves to align teacher expectations with actual learning progress, ensuring teachers confidently recognize productive learning even when behavioral success is not yet visible.}


This conceptual framework must be operationalized into a computable reinforcement learning metric. Specifically, the mathematical definition of Insight must be \textit{scalable}, i.e. capable of capturing both sudden step-insights and prolonged, multi-step episode insights across different time frames, and \textit{generalizable}, ensuring the signal is grounded in universal RL mechanics rather than hard-coded to a specific environment.

\subsection{Faithful, Scalable and General Formalization of ``Insight"}
To fulfill the Faithfulness, Scalability and Generalizability criteria, the mathematical formalization of Insight must be independent of specific task parameters or algorithmic architectures. 
The misalignment between teacher mental model and agent learning process corresponds to the evaluative dimension of observation integration and value updating. 
Therefore, Insight is in this work derived from the Temporal Difference (TD) Error \cite{broekensTemporalDifferenceReinforcement2018}, a universal metric in RL that inferently quantifies the magnitude of an internal evaluative update. 
Because the human teacher needs to understand the amount of learning rather than the direction of the value change (since both a pleasant surprise and a massive disappointment represent equal amounts of ``learning insight"), the absolute value of the error is used.
To ensure the signal scales across different task granularities and algorithms and does not manifest as a series of easy-to-miss instantaneous ``blips",  
we formalize Insight as a moving average over the absolute TD error within a given temporal scope (a window\footnote{Window size N should be calibrated to the task's temporal scale: while N=5 may suffice for discrete grid worlds, complex continuous tasks require larger windows to ensure the signal aligns with human-perceptible behavioral segments. Optimal N for specific settings should be determined via human-centered expression studies and is outside the scope of our analysis.} of size N):

\vspace{-10pt}
\[
\mathrm{Insight}_t = \frac{1}{N}\sum_{i=1}^{N} \left| \delta_{t-i} \right|
\]
\vspace{-10pt}

with the temporal-difference error defined as usual:

\vspace{-10pt}
\[
\delta_t = r_t + \gamma\, Q(s_{t+1}, a_{t+1}) - Q(s_t, a_t)
\]
\vspace{-15pt}

Just as affect serves as an internal evaluative currency for biological organisms, value-based mechanisms (such as TD error) act as the native evaluative substrate in RL. This operationalization ensures an algorithmically faithful and generalizable yet human interpretable reflection of the agent's actual learning progress.

\section{Evaluation Methodology}

\subsection{Task and Simulation Environment}
\begin{figure}
    \centering
    \includegraphics[width=0.6\columnwidth]{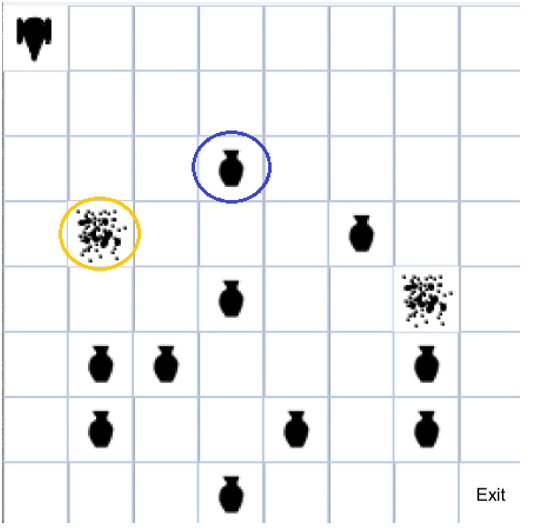} 
    \caption{$8 \times 8$ grid-world with Start, Exit (+1.0), Dirt (+0.5) and Vases (-1.0).}
    \label{fig:Task}
    \vspace{-17pt}
\end{figure}

To evaluate the feasibility of generating the Insight signal, a controlled environment where internal state changes can be mapped directly to observable task events is required. 
Further the task should leave enough room for teacher interventions as well as teacher misinterpretations during the learning process. 
To most accurately represent the settings under which the original, empirically motivated mental model misalignments was found, the cleaning robot task from \cite{hilpertCanYouSee2025} was reproduced:
The agent operates on an $8\times8$ grid (64 cell locations) containing $M$ dynamic dirt patches. To maintain high transparency, a flattened tabular state space was utilized, where a full state $s$ combines the robot's spatial index and a bitvector of cleaned dirt patches (see Figure \ref{fig:Task}). 
For $M=2$, this yields a finite state space of size 256.
To create complex credit-assignment challenges, the environment features sparse positive and negative rewards: +0.5 for cleaning a dirt patch (1x/episode and dirt), -1.0 for breaking a fixed set of obstacles (vases), and +1.0 for reaching the terminal exit state. For the agent, a standard tabular Q-learning algorithm with TD update and e-greedy exploration policy was used $(\alpha = 0.1, \gamma = 0.95, \epsilon_{initial} = 0.1)$, to ensure the TD updates are directly observable, serving as a transparent testbed for the proposed Insight signal. For the temporal scope, a window of N=5 was chosen.
To ensure robustness, results are averaged over five independent seeds, with each simulation run consisting of 300 episodes to observe the signal's evolution over the entire learning process.
Following recommendations for open research practices \cite{wessler2021empirical}, all materials derived for this experiment (e.g., relevance analysis, modular simulation framework, and evaluation protocols) are openly available on OSF\footnote{\url{https://osf.io/wyjrv/overview?view_only=9d389029713d46c7a191589e8b529f16}}.

\subsection{Evaluation Framework}

To evaluate the efficacy of the Insight signal as an algorithm-faithful yet human-understandable substrate, a multi-criterion validation focused on three core properties was conducted:

\begin{enumerate}
    \item \textbf{Spatiotemporal Fidelity (Micro Dynamics):} Assesses if the signal triggers precisely during internal value updates, measuring the temporal/location correspondence between signal magnitude and reward-related events, ensuring spikes occur only during impactful learning (not generating false positives during action repetition). 
    \item \textbf{Evolutionary Stability (Macro Dynamics):} Assesses the signal's trajectory over the agent's lifetime, evaluating the relationship between Insight magnitude and cumulative returns, verifying that the signal decays as the policy stabilizes and serves to understanding the structure and patterns of the signal.
    \item \textbf{Evaluative Saliency (Event-specific Contrasts):} Assesses signal responsiveness to different stimuli, using mean magnitude comparisons across event types to ensure Insight distinguishes the importance of varied interactions (e.g. finding dirt vs. crashing into a vase).
\end{enumerate}

\section{Results}
The Insight signal was examined in a multi-criterion evaluation process. While various hyperparameter configurations (i.e. variations in learning rate, epsilon, epsilon decay) were evaluated during the experiments, for conciseness only one configuration is reported here; the complete set of extended results and alternative plots is available in the OSF repository.

\begin{figure}
    \centering
    \includegraphics[width=\columnwidth]{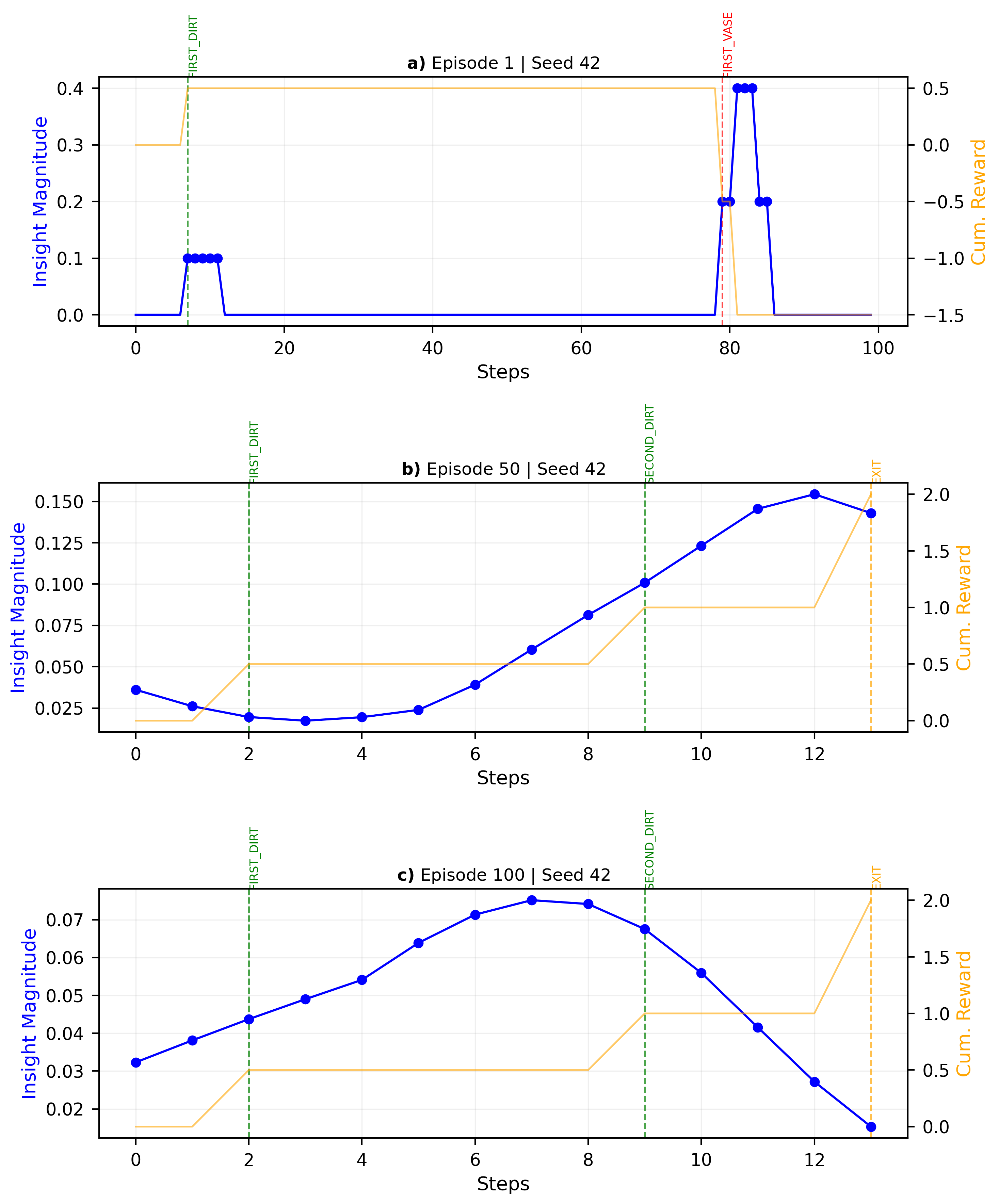} 
    \vspace{-23pt}
    \caption{Micro Dynamics: Insight Traces (3 representative episodes): (a) Discovery: Reactive signal peaking at rewards; (b) Transition: Continuous Insight Signal despite behavioral convergence with a peak towards the Exit tile; (c) Beginning Mastery: Peak moves towards initial steps.}
    \label{fig:insighttraces}
    \vspace{-5pt}
\end{figure}

\begin{figure}
    \vspace{-5pt}
    \centering
    \includegraphics[width=\columnwidth]{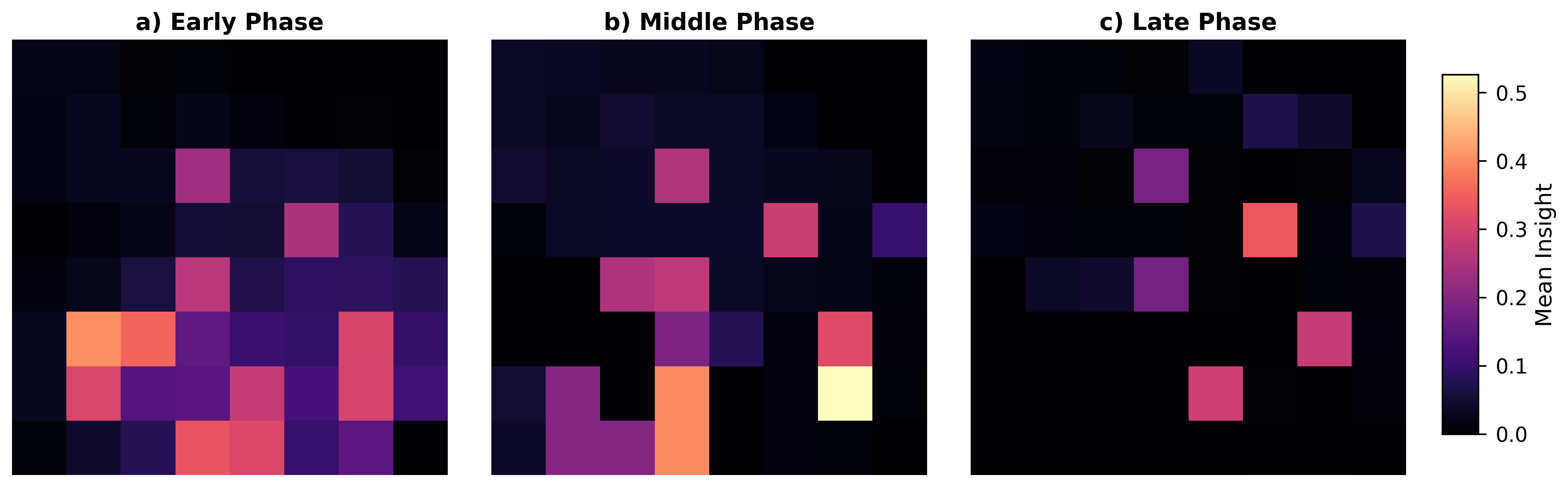} 
    \vspace{-23pt}
    \caption{Mean Insight heatmaps (5 seeds, 100 episodes per pahse). Early phase shows diffuse exploration; Middle shows localized saliency at obstacles; Late shows signal decay upon policy stabilization except for trauma locations}
    \label{fig:heatmaps}
    \vspace{-15pt}
\end{figure}

Insight was designed to reflect meaningful learning progress, indicating moments in which new experiences produce substantive changes to the agent's evaluative predictions. Therefore the first step was to examine whether Insight corresponds to reward-related events from which the agent learns.
Results show that at the start of learning, Insight temporally corresponds in magnitude to both positive and negative rewards (Fig \ref{fig:insighttraces} a): every reward-related event is followed by an immediate insight spike proportional in magnitude to cumulative absolute rewards. Additionally, Insight spikes happen at or around reward locations (Fig \ref{fig:heatmaps} a).

\begin{figure*}
    \centering
    \includegraphics[width=\textwidth]{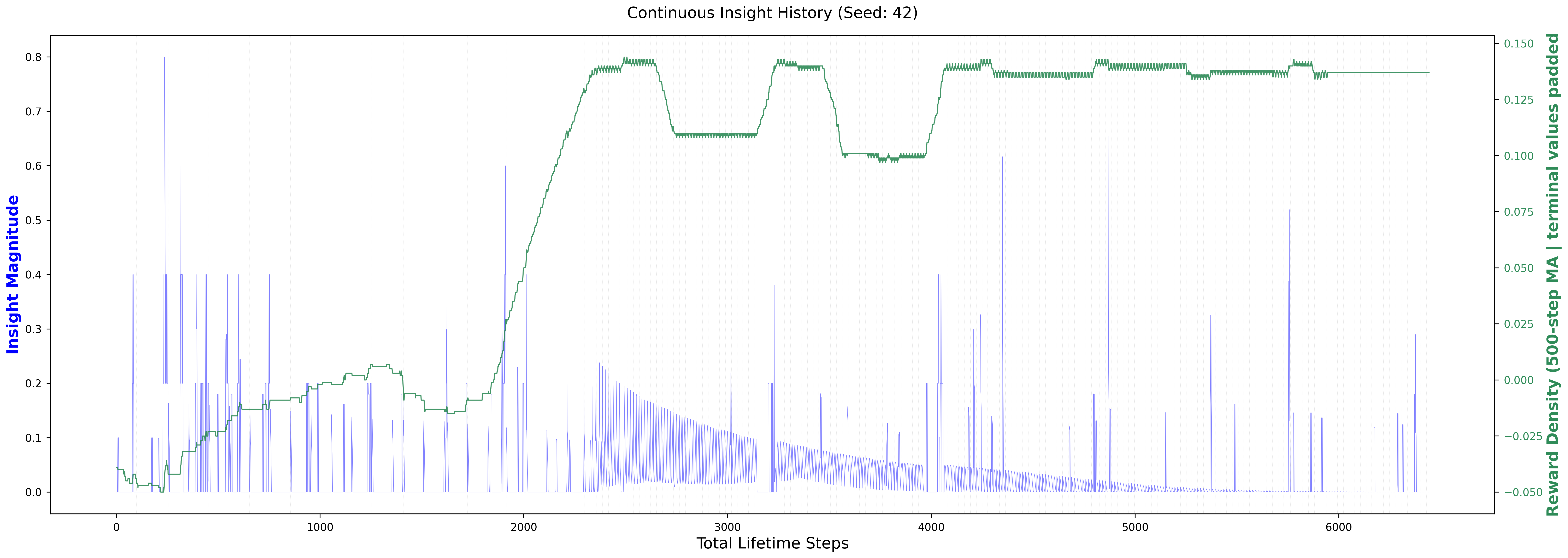} 
    \caption{Full Insight history. Blue trace indicates Insight; green line denotes reward density. Note the signal's heavy peaking during early exploration followed by a systematic pattern with sparse peaks during exploitation.}
    \label{fig:fullhistory}
    \vspace{-20pt}
\end{figure*}

\begin{figure}
    \centering
    \includegraphics[width=\columnwidth]{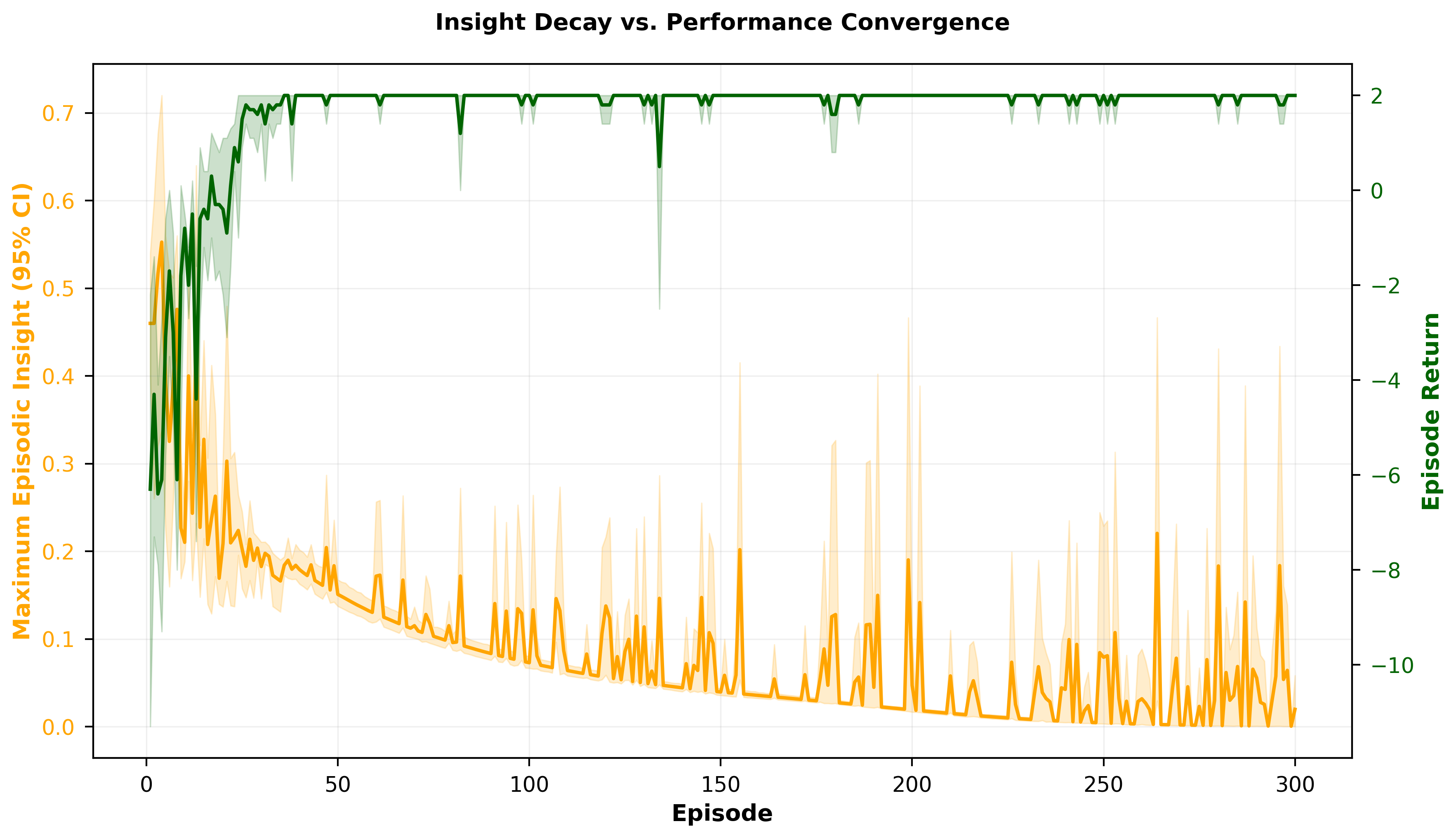} 
    \caption{Macro-dynamics (5 seeds). Max Insight (orange) decays as Episode Return (green) plateaus, demonstrating the signal's sensitivity to performance convergence. Shaded areas represent 95\% confidence intervals across all seeds.}
    \label{fig:Convergence}
    \vspace{-15pt
    }
\end{figure}

Further, for behaviorally converged traces (i.e. traces in which no vases are hit anymore and all three rewards collected in a fixed number of steps) the signal follows the backpropagation of values. Results show that Insight is non-zero at non-event steps (see Fig \ref{fig:insighttraces} b and c), reflecting that the agent is also learning from steps it is not directly rewarded for while it is still converging. 
Additionally, the Insight ``peak" in these traces can be seen to propagate back through the policy. Further tests (not reported here) revealed that this Insight peak corresponds to the learning rate in the update mechanism.
This moving peak therefore makes the actual state of convergence transparent, corresponding to ongoing backpropagation of values. 
The spatial distribution of Insight in the middle phase shows localized saliency at task-relevant obstacles rather than following a singular optimal path (Fig \ref{fig:heatmaps} b). By the late phase, Insight effectively extinguishes across most of the state space as the value function stabilizes. However, high saliency persists at high-penalty locations, where their avoidance prevents the signal from habituating (see Fig \ref{fig:heatmaps} c).
When looking at the full history of steps taken by the agent as compared to reward density (see Fig. \ref{fig:fullhistory}), two main findings can be observed. 
First, high Insight peaks occur within the first 2400 steps, indicating significant internal updates, before any visible increase in reward density. Once reward density plateaus (step 2500 and onward), it remains stable despite occasional exploratory blips. Crucially, the Insight signal persists in systematic, decaying waves during this phase. This demonstrates that while the external policy has converged, the signal continues to reflect ongoing learning and backpropagation during exploitation. 
A more aggregated view per episode (see Figure \ref{fig:Convergence}) shows the same trend consolidated on a macro level: A volatile increasing return in the exploration phase with spikey per episode maximum Insight signals (as a proxy for an observable affective signal) as well as a steady return and the expected decaying Insight magnitude (with the occasional exploration-driven hiccup due to non-zero epsilon) in the exploitation phase.
Additional experimentation omitted in the paper as expected produced a steeper decay in the insight with increased learning rates and a longer exploration phase and more ``hiccup" spikes with smaller epsilon decay rates.  

\begin{figure}
    \centering
    \includegraphics[width=\columnwidth]{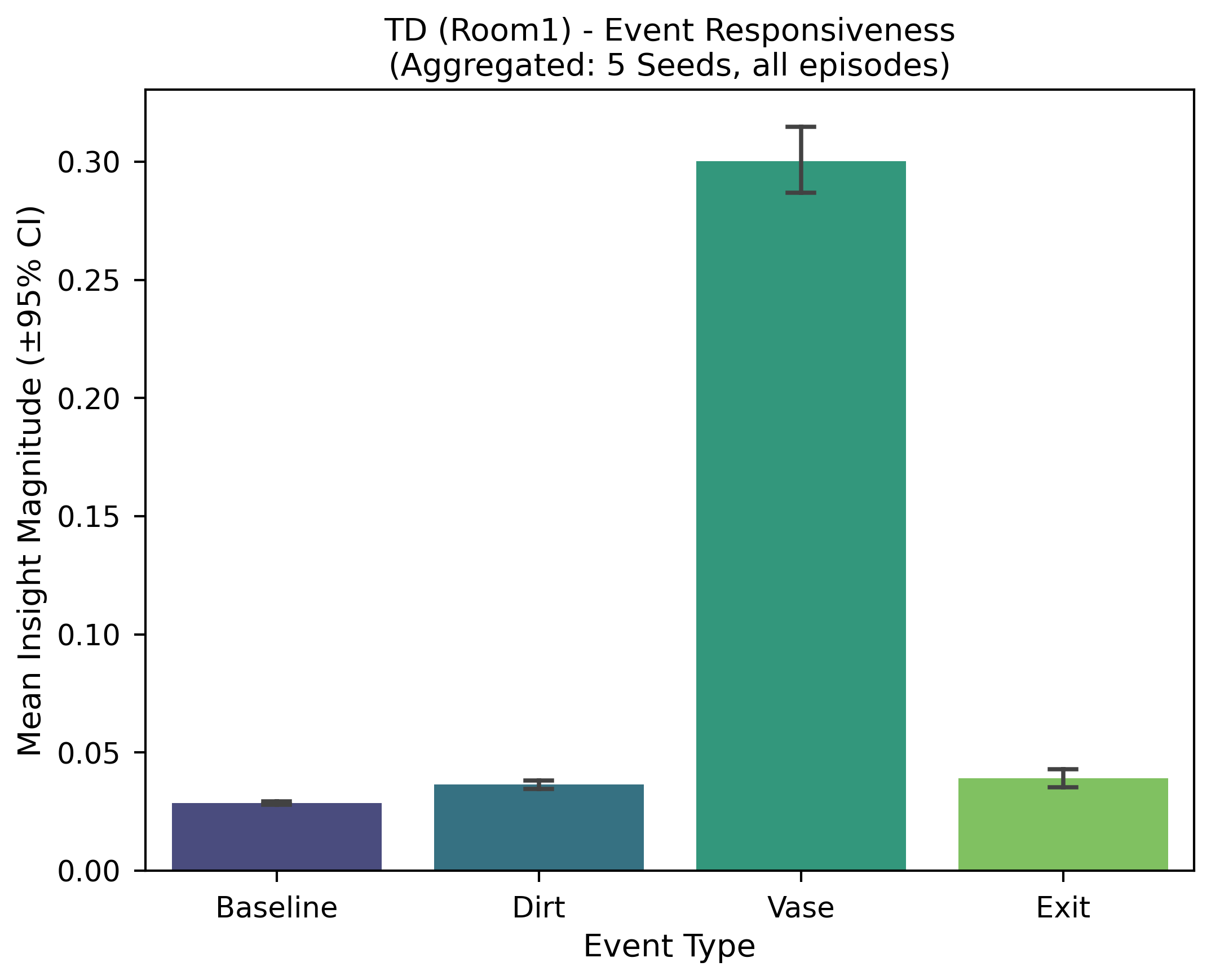} 
    \caption{Mean Insight magnitude effectively distinguishes between neutral navigation (Baseline) and salient events (Dirt, Vase, Exit). High-magnitude responses to negative rewards quantify the ``trauma-like'' sensitivity of Insight.}
    \label{fig:contrastoverall}
    \vspace{-15pt}
\end{figure}

When looking at the Insight signal by Event type, it can be observed that even floor tiles from which the agent does not receive any rewards generate an Insight signal as values are backpropagated and contribute to learning. Further, while dirt patches and exit tiles do generate distinguishable insight signals, the Insight from stepping on a vase (i.e. negative rewarded) tile, produces by far the highest average Insight magnitude (see Fig. \ref{fig:contrastoverall}), which is expected as negative events are avoided and thus the average insight signal cannot decay.

\section{Discussion}

In this work, Insight was defined, formalized, implemented and subjected to a multi-criterion validation to evaluate its efficacy in reflecting meaningful learning progress as a novel RL-grounded artificial affect and explainability signal.

The results show that Insight is a functional proxy for the agent's internal evaluation process, rather than a simple reward log. 
This is evidenced by the spatio-temporal correspondence between Insight magnitude and both positive and negative reward events.
By scaling according to the underlying reward structure, Insight translates raw scalar feedback into internal evaluative currency.
By utilizing absolute TD-magnitude and temporal smoothing, Insight transforms fleeting, directional spikes into a sustained signal of learning salience. This ensures that both surprising successes and failures are captured as informationally substantive updates. This signal serves as an algorithmic substrate for `Aha!' moments, indicating events a meaningful shift the agent's future expected return.
Beyond immediate reward-related events, the observed wandering maximum signal peak from the terminal reward location back toward the start as learning progresses, provides a direct reflection of agent learning mechanics, specifically the learning rate, indicating the velocity of the credit assignment.
This provides technical transparency that goes beyond event-tracking: an observable proxy for the speed of integration of new experiences into its long-term policy.
In that sense, Insight successfully reflects that the agent is assigning a meaning to and learning from the states leading up to a goal, making it distinct from pure reward logging. 
The distinction between behavioral convergence and representational stability is made transparent by Insight.
An external observer might perceive a behaviorally converged agent as having `finished' learning following an optimal path (Fig \ref{fig:heatmaps} and \ref{fig:insighttraces} b, also \ref{fig:fullhistory}), but continued Insight signals (also at non-event tiles) reveals behaviorally invisible ongoing active value propagation and internal model refinement, making transparent that the agent is still learning. 
In this way, Insight disambiguates between active consolidation and habitual execution. 
As the agent approaches mastery, the Insight signal undergoes a `spatial refinement', firing only during rare off-policy explorations. 
For a teacher, this transforms the signal from a `learning progress' bar into a `surprise detector' and makes the invisible process of stabilization observable, directly addressing the mental model misalignment of ``Absence of learning". 
Since Insight is derived from the TD Error, 
it functions as a premonitory cue, flagging internal reorganization predictively before it manifests as a change in the agent's behavior. This transforms the human's role from reactive to proactive: a teacher can anticipate shifts in strategy. 
However, whether or not teachers intuitively understand this link needs to be tested in an experimental setting in the future.
As it is grounded in the TD error, Insight is naturally influenced by the learning rate, reflecting credit assignment and value propagation velocity. 
Such a signal offers significant potential for behavioral diagnostics in complex agents where direct code access is unavailable: 
an observer can diagnose whether a robot is ``stuck", ``consolidating", or ``malfunctioning" simply by interpreting its affective output.
Further, it serves as an indicator for adaptivity and learning dynamics of a certain algorithm, helping prevent over- or under-convergence when agents transition to new environments.
It also captures the ``hidden" representational shifts that must occur before the agent can successfully navigate to a reward.
On a macro-dynamic scale, the inverse relationship between Insight magnitude and learning progress remains stable throughout training. As cumulative Insight systematically decreases alongside increasing returns, it provides a longitudinal indicator of the agent’s transition from exploration to stable exploitation. 
This suggests that Insight is also a robust metric for internal stabilization rather than a byproduct of transient rewards.  For teachers, this decay serves as a functional teaching trigger: when magnitude drops and the behavior is desired, the utility of further instruction is limited, however, when magnitude drops and the behavior is \emph{undesired}, instruction is needed. 
By empowering teachers to their own subjective stopping points, Insight addresses the `Absence of Learning' misalignment. Effect on teaching behavior needs future experimental investigation.
The event-related analysis reveals that Insight goes beyond mirroring reward activity, but rather disambiguates the evaluative impact of different environmental interactions of differing saliency. 
First, signal presence on ``neutral" floor tiles indicates an `invested' agent updating its trajectory valuation, rather than a `reactive' agent that is `just moving` learning only upon reward contact.
Second, Insight magnitude effectively differentiates between the relative weights of different goals and their visitation. While dirt patches and exits generate distinguishable Insight peaks, 
the most striking finding is the `Vases' (negative rewards) producing by far the highest Insight magnitude.
This disproportionate response can be interpreted as a functional `trauma' pattern. Unlike positive rewards which decay through repeated visitation, rare negative encounters never habituate. 
Each encounter remains as a `shocking' event in the agent's learning history. 
As confirmed by late-phase heatmaps, these `hot' trauma signatures provide permanent markers of environmental danger, visible to the teacher even after the agent learns to avoid the obstacle.
This suggests that the agent `remembers' these negative encounters as significantly more traumatic than positive ones. 
This non-habituating cue identifies critical failure points and offers diagnostic potential without direct code access. 
While `trauma' should not be taken literally, it can serve as an anthropomorphized metaphor that may foster intuitive understanding in human teachers (similar to \cite{salichs2011new}).
For example, a robot mysteriously avoiding a clear corner could have its behavior traced to a `trauma spike', perhaps an encounter with a now-removed obstacle. 
Insight thus moves beyond simple transparency, serving as a functional stethoscope for the internal health and history of the RL process.


While formalized here as an affective signal, Insight is, at its core, the Temporal Difference signal packaged for human interpretability. The results of the simulation study suggest that this fundamental metric of RL is not only essential for optimization but is also a potent vehicle for explainability \cite{broekensTransparentRobotLearning2019}. Further, this formalization of Insight conceptually mirrors the human `Aha!' experience \cite{kounios2009aha}.  We take a ``less is more" approach to explainability; the flatter the model and the more direct the signal, the more faithfully it reflects the agent's internal state. 
By using such algorithm-faithful signals as explainability mechanism, transparency becomes an inherent property of the learning process itself, effectively realigning the human-robot loop. 
We do not claim equivalence between the TD error and affect or human insight. Rather, this work provides the theoretical foundation to test whether artificial affect signals 
can be designed, formalized and simulated in order to serve as an anthropomorphized inference canvas that in the future may be used to improve he human teacher's understanding of RL agents' learning processes.

\section{Limitations and Future Work}
While this first instantiation of insight as an affective signal used as an explainability mechanism for HIRL follows a principled approach of design criteria and evaluation, several open factors remain unexplored in this paper. 
First, Insight is only one of several signals that could close the ``Absence of Learning" misalignment. different teacher expertise levels may require alternative affective substrates. 
Second, while an update rule-based signal aligns with the identified misalignment, it represents a preliminary structural proxy. Future formalizations could more deeply integrate neuro-scientific models of the “Aha!” experience \cite{kounios2009aha} to further enhance interpretability. Methodologically, the N=5 window was manually calibrated for the grid-world task. As grid-worlds offer a simplified testbed, future work must evaluate the signal's generalizability to high-dimensional or dense-reward tasks (e.g., robotic manipulation), including adaptive temporal scaling to maintain signal saliency across complex behavior.

\section{Conclusion}

This work presented ``Insight", an algorithm-faithful affective signal designed as an explainability mechanism to enhance transparency in Human-Interactive Robot Learning (HIRL). By mapping the TD error (a fundamental reinforcement learning metric) to a human-interpretable, teaching-relevant concept, it leverages the natural dual purpose of affect to support intuitive interpretability. The multi-criterion validation confirmed that Insight is not a mere mirror of reward activity, but a rich evaluative currency reflecting meaningful learning events, the learning phase and the often invisible backpropagation of learning. The core contribution of this paper lies in demonstrating that highly technical RL dynamics can be effectively packaged into intuitive affective abstractions. We have proposed a principled way for designing and computationally modeling such signals. This way, this work lays the foundation for a broad class of teaching-relevant, algorithmically faithful affective signals capable of closing the loop between human teaching intuition and robot learning processes.

\section*{Ethical Impact Statement}

\subsection*{Societal Impact and Applications}
This work introduces ``Insight,'' a transparency mechanism designed to align human teacher expectations with the internal learning progress of Reinforcement Learning (RL) agents. By translating technical metrics (TD-error) into human-interpretable affective signals, this research has the potential to make Human-Interactive Robot Learning (HIRL) more accessible to non-experts. This could democratize the training of domestic and assistive robots, allowing lay users to effectively guide AI behavior without requiring deep algorithmic knowledge.

\subsection*{Potential Risks and Mitigation}
While the ``Insight'' signal is intended to foster trust and efficiency, it carries risks associated with anthropomorphism and over-reliance:
\begin{itemize}
    \item \textbf{Deception and Over-trust:} Humans naturally attribute human-like intent to affective cues. There is a risk that users may over-rely on these ``simulated'' emotions, perceiving the robot as more sentient or capable than it truly is. To mitigate this, our framework emphasizes \textit{Faithfulness}, ensuring the signal is mathematically grounded in actual learning updates ($|\delta_t|$) rather than being a superficial ``black box'' display.
    \item \textbf{Manipulation:} Affective substrates could theoretically be used to manipulate human teachers into providing more feedback through signals that mimic distress or satisfaction. Our design criteria specifically focus on \textit{Relevance (Teaching Focus)} to ensure signals serve instructional transparency rather than emotional manipulation.
    \item \textbf{Bias in Interpretation:} The interpretation of the ``Aha!'' moment metaphor may vary across different cultures or neurodiverse populations. Future work must involve diverse user studies to ensure the \textit{Interpretability} criterion holds across a broad demographic spectrum.
\end{itemize}

\subsection*{Data and Privacy}
The empirical grounding for this work utilized an existing, openly accessible dataset containing qualitative human inferences. No new human subject data was collected for the simulation studies presented here. However, we acknowledge that implementing this system in real-world settings would require careful handling of teacher-learner interaction data to protect user privacy.

\subsection*{Research Integrity and Openness}
To support reproducibility and transparency, all simulation environments, algorithmic implementations, and evaluation protocols have been made available via an anonymized repository. This allows the community to audit the ``Insight'' signal's behavior and ensures the work can be independently validated.

\section*{Acknowledgments}
This research is partly sponsored by the Hybrid Intelligence project, grant number 024.004.022. Special thanks to Thomas Moerland, Anna Lea Reinwarth, Hartmut Hilpert and Albert Roca Saura.

\bibliographystyle{IEEEtran}
\bibliography{affsim_acii_CR}

\end{document}